\documentclass[a4paper, amsfonts, amssymb, amsmath, reprint, showkeys, nofootinbib, twoside]{revtex4-1}
\pdfoutput = 1
\usepackage[english]{babel}
\usepackage[utf8]{inputenc}
\usepackage[colorinlistoftodos, color=green!40, prependcaption]{todonotes}
\usepackage{multirow}
\usepackage{float}

\usepackage{amsthm}
\usepackage{mathtools}
\usepackage{physics}
\usepackage{xcolor}
\usepackage{graphicx}
\usepackage[left=18mm,right=13mm,top=35mm,columnsep=15pt]{geometry} 
\usepackage{adjustbox}
\usepackage{placeins}
\usepackage[T1]{fontenc}
\usepackage{lipsum}
\usepackage{csquotes}

\usepackage[normalem]{ulem}

\usepackage[pdftex, pdftitle={Article}, pdfauthor={Author}]{hyperref}
\begin{document}

\title{Designing interferometers within a single optical beam}

\author{Bereneice Sephton, Rakhi Thomas, Carlo Schiano, Francesco Reda, I Komang Januariyasa, Filippo Cardano, Bruno Piccirillo, Marcella Salvatore, Stefano Luigi Oscurato, Corrado de Lisio and Vincenzo D'Ambrosio}
    \email[Correspondence email address: ]{vincenzo.dambrosio@unina.it}
    \affiliation{Dipartimento di Fisica "E. Pancini", University of Naples Federico II, Naples, Italy}

\begin{abstract}
 Interferometry provides highly sensitive access to optical phase and is central to much of modern metrology and phase imaging methods. Conventional implementations, however, often face trade-offs between mechanical stability and experimental or computational complexity.
 Here, we present a general framework for designing custom interferometers within a single optical beam by exploiting structured light.  This approach yields compact, robust common-path configurations that bypass the need for complex post-processing and can easily be  integrated into existing setups. 
  We demonstrate the versatility of this concept by designing a range of interferometers, each tailored by the structured mode, and implement them through active and passive modal conversion optics, proving its adaptability to different experimental requirements. To showcase the practical utility of our framework, we apply it to quantitative phase imaging over a variety of physical samples, showing excellent agreement with atomic force microscopy benchmarks. 
  Furthermore, we emphasise the flexibility of our structured light interferometers by mapping phase objects to a choice of either amplitude or polarisation, the latter providing a direct route toward real-time phase-retrieval. This cost-effective approach offers a practical, high-throughput solution for phase-sensitive metrology across fields such as fundamental physics, biology, and material science.

\end{abstract}

\maketitle

 Purposefully tailoring the amplitude and phase of light \cite{forbes2021structured,he2022towards} has wide-reaching impact, facilitating applications from optical trapping \cite{yang2021optical} to communications \cite{willner2021orbital}. Alternatively, retrieving the phase structure of light is less direct as the inherent phase insensitivity of conventional image sensors necessitates indirect approaches, such as interferometry, where a reference beam is mixed with the signal beam \cite{ichioka1972direct,frantz1979optical,van2022optical,cui2022quantitative,besse2012spatially}. This approach provides highly sensitive, non-contact phase measurements and finds use in surface metrology \cite{zhao2025review}, precision displacement sensing \cite{huang2024review}, and optical coherence tomography \cite{huang1991optical,bouma2022optical}.
 In particular, traversing samples can reveal important information such as topography \cite{de2015principles} and composition, essential in biological, medical \cite{park2018quantitative} and material applications \cite{fang2017nanomanufacturing,brand1995calibration,dai2005accurate}, giving way to the thriving field of quantitative phase imaging \cite{nguyen2022quantitative,chaumet2024quantitative,huang2024quantitative}.  Unfortunately, light beams propagating along spatially separated paths usually result in instability \cite{shaked2012quantitative,bai2015common,rodriguez2024axis}, requiring either active correction \cite{krishnamachari2006active,hacker2023phase} or periodic realignment, and generally need bulky frameworks, which are not simple to integrate into existing systems.

This has inspired a myriad of alternative strategies \cite{park2018quantitative,mir2012quantitative,nguyen2022quantitative}, which often compromise between experimental and computational complexity. Moving away from traditional interferometry removes the instability associated with a reference arm, but typically requires significant computational overhead \cite{streibl1984phase,paganin1998noninterferometric,barone2002quantitative,tian20153d,katkovnik2017computational,wu2021wavelength,garcia2006digital,gao2023iterative,sun2023compressive}. Segmented approaches, such as Shack–Hartmann wavefront sensors \cite{gong2017optical}, avoid interferometric instability but at the cost of reduced resolution \cite{mir2012quantitative}.
Alternatively, common-path interferometry seeks to provide stability with reduced computational complexity by having both beams travel along the same or very similar paths \cite{nguyen2022quantitative}. For instance, double-passing schemes \cite{dyson1957common,dyson1963very,bvehal2019quantitative,wei2020w} minimally separate the reference and send it back in the same path, shear interferometry \cite{schreiber1997lateral,bates1947wavefront,murty1970compact,aleman2019shearing} replicates and interferes the beam with itself, and self-referencing interferometry \cite{singh2012lateral,lee2014quantitative,bvehal2019quantitative,sardana2023dielectric} places the sample in only a part of a duplicated beam so that the structured areas do not overlap. Choice of each approach often relies on respectively reconciling between varying trade-offs. Namely, bulky and complex experimental configurations, additional extraction calculations or resolution limitations with sample size to reference overlap constraints.

  \begin{figure*}
\centering
\includegraphics[width=1\linewidth]{Fig1_new.png}
\centering
\caption{\textbf{Structured light as an interferometer.} (a) Traditional interferometers use elements such as beam-splitters (BS) to split light into two paths (1 and 2) before recombining them to observe any phase differences that now modulate the intensity of the exiting light. (b) Light can similarly be split (forward arrow) into two separate collinear optical modes ($\Phi_1$ and $\Phi_2$) using phase-modulating optical elements which we refer to as mode-splitters (MS). Performing the reciprocal operation (reverse arrow) recombines the light back into the same optical mode, forming a structured light interferometer. Insets (a,b) highlight these splitting operations with the modes in (b) shown after propagation for conceptual clarity. (c) Structured light interferometers can be used to perform quantitative phase imaging where modes with vortex, Bessel or displaced structures, for example, become a reference (Ref.) which propagates alongside another Gaussian mode (Pr.) that probes an object (Obj.) before being recombined by reversing the generation process. Inset (i) illustrates how spatial separation between the structured vortex and unstructured Gaussian components allows only the probe to interact with the object of interest. By using polarisation-sensitive structuring elements as MS we selectively mark and revert the structured mode for interference upon recombination. Recombination operations either (d) return the reference polarisation to that of probe and map the object phase directly into intensity or (e) retain the orthogonal polarisation and map the object phase to the polarisation structure instead. A $\pi$-step phase object is shown to illustrate the effect. Arrows on the intensity distributions represent local polarisation states.}
\label{Fig1}
\end{figure*}

In this paper, we introduce a common-path approach that uses structured light modes to form a collinear interferometer, which we term a structured light interferometer (SLI). This provides a compact and robust single-beam geometry that offers a versatile route to interferometric design with minimal computational overhead. 
Here, structuring the optical field splits it into co-propagating probe and reference modes that act as conventional interferometer arms.
Reversing this transformation recombines these modal arms into a single, interfering spatial mode.
This common-path nature inherently suppresses environmental phase noise, providing improved stability compared to conventional interferometers.
We illustrate the generality of our approach using three different modal families and across both passive and dynamic technologies, each giving rise to a tailored interferometric design in a quantitative phase imaging setting. 
Furthermore, we show that phase information can be mapped to either the intensity or the polarisation of the output beam. In each case, we obtain excellent reconstruction of phase samples with different geometries. Together, these results establish a robust and flexible approach for phase-sensitive metrology with minimal post-processing. This provides a practical route to non-contact, real-time measurements across biology, materials science and industry that are easily integrated and can be tailored for each application.

\section*{Results}
\noindent \textbf{Structured light interferometers.} In traditional interferometry, light is split into different paths and then recombined, often using beam splitters (Fig. \ref{Fig1} (a)). Inspired by this principle, we exploit modulation optics as mode splitters (MSs) to divide the optical field into two spatially distinct modes and subsequently recombine them back into a single one. Figure \ref{Fig1} (b,c) illustrates our concept where the interferometer is formed between generation and recombination MSs. The first MS imparts structure to the optical field traversing it, while the second performs the inverse transformation. In between, the co-propagating modes separate spatially to form the individual arms of the SLI. Here, full control relies on addressing each mode independently.
By choosing polarisation as the discriminating degree of freedom, mode-selective operations are possible with readily available optical elements such as metasurfaces \cite{devlin2017spin,devlin2017arbitrary}, spatial light modulators \cite{forbes2016creation}, and liquid crystal-based geometric phase devices \cite{piccirillo2010photon} which facilitate modulation (or lack thereof) based on the polarisation state of the incident light. This dual encoding strategy further highlights the flexibility of the SLI framework, allowing one to choose between intensity-based and polarisation-based phase retrieval.

\noindent \textbf{Quantitative phase imaging with SLIs.} As a practical demonstration of our concept, we describe its application to quantitative phase imaging (Fig. \ref{Fig1} (c)). 
We start with an input optical field having a transverse Gaussian spatial profile, $G(r) = \mathcal{N}e^{-(r/w_0)^2}$, and right circular polarisation (RCP, $\hat{e}_R$). Polar coordinates in the transverse plane are ($r$, $\phi$), while $\mathcal{N}$ is a normalisation term and $w_0$ the beam waist.
For convenience, the input mode is transformed into a balanced superposition of the original field and a component modulated by the phase profile $\Phi(r,\phi)$, allowing the Gaussian mode to serve as one interferometric arm with the advantage of having a flat phase profile ($\Phi = 0$).
This transformation can be implemented by a  geometric phase-based device tuned to 50\% efficiency (see SI for details), 
\begin{equation}
    G(r)\hat{e}_R \rightarrow \frac{1}{\sqrt{2}}\left[G(r)e^{i\Delta\theta}e^{i\Phi(r,\phi)}\hat{e}_L + G(r)\hat{e}_R\right],
    \label{Eq_split}
\end{equation}
where the modulated component undergoes a flip from right to left circular polarisation (LCP, $\hat{e}_L$). Here, $\Delta\theta$ represents a constant, arbitrary intermodal phase offset.  Note that Eq. \ref{Eq_split} is given for the plane of the modulation optic and we neglect the propagation terms, considering all operations at the beam-waist (i.e. the image and focal planes of the generation optic).  
Vortex modes, characterised by a ring-shaped intensity profile, are used as illustrative examples in Fig.~\ref{Fig1} (c–e), which can be generated with a \textit{q}-plate \cite{slussarenko2011tunable} of charge \textit{q}, imparting $\Phi(r,\phi)=2q\phi$. Spatial separation arises upon propagation and is maximised in the far field, realised with a 2$f$ lens configuration (Fig.~\ref{Fig1}(c)). A doughnut-shaped distribution then forms in the focal plane, approximately $\mathcal{F}[G(r)e^{i 2q\phi}]$ , where $\mathcal{F}[\cdot]$ represents the Fourier transform and the uncollected light is neglected. With the reference mode sufficiently separated in the transverse plane (as in inset Fig. \ref{Fig1} (c,i)), only the unconverted Gaussian mode (Pr.) interacts with the phase object and carries the related information. This can be seen in the transverse spatial field of the interferometer ($\Psi$) in the object plane,
\begin{equation}
\begin{split}
 \Psi(r,\phi) = \frac{1}{\sqrt{2}}(&\mathcal{F}[G(r)e^{i\Delta\theta}e^{i\Phi(r,\phi)}]\hat{e}_L \\
 &+ e^{i Obj(x,y)}\mathcal{F}[G(r)]\hat{e}_R),
 \end{split}
\end{equation}
\noindent where we have introduced the phase retardation, $Obj(x,y)$, originating from the object. By reverting the reference mode with an identical MS in the image plane of the generation optic, 
$G(r)e^{i\Delta\theta}e^{i\Phi}\hat{e}_L \rightarrow G(r)e^{i\Delta\theta}e^{i\Phi}\hat{e}_R\times e^{-i\Phi} = e^{i\Delta\theta}G(r)\hat{e}_R$, the optical field is easily recombined into the same mode.

\begin{figure}
\centering
\includegraphics[width=1\linewidth]{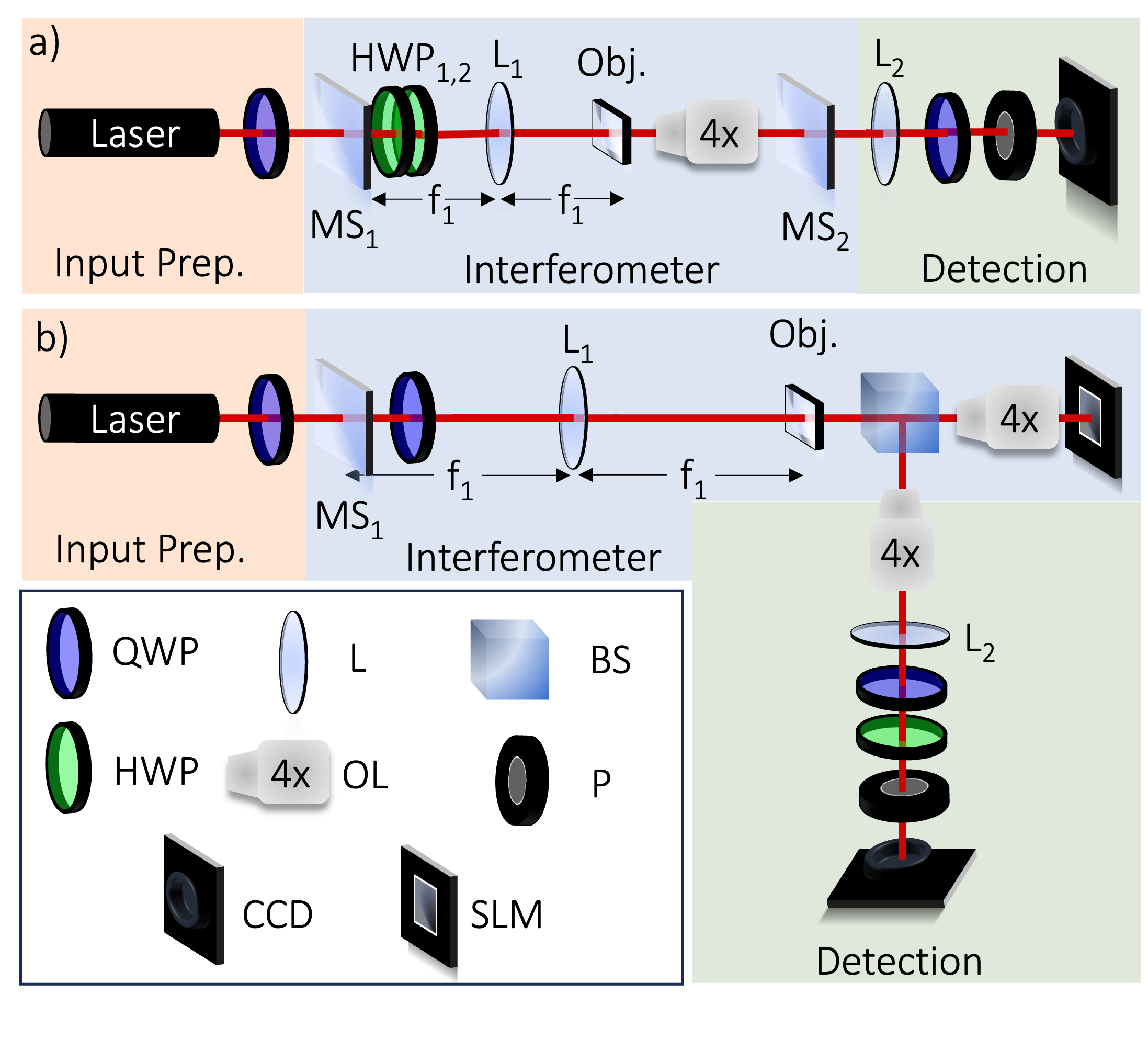}
\centering
\caption{\textbf{Experimental setup.} Experimental setups used for quantitative phase imaging with SLIs. A horizontally polarised laser with a wavelength of $\lambda$ = 803 nm was converted to RCP with a quarter wave plate (QWP), before traversing two optical devices acting as MSs. Each MS was imaged onto the other with a 4-$f$ telescope (lens, $L_1$, and a 4x objective lens, OL). All objects were placed in the Fourier plane of $L_1$ and imaged onto the CCD with additional lens ($L_2$) after the OL. The interferometer is set in transmission in (a) with tuneable geometric-phase based optical element being used for both MSs. A QWP and polarizer (P) selected the RCP component of the light exiting the interferometer. Two half waveplates, $HWP_{1,2}$ within the interferometer generated a phase difference between the reference and probe based on the relative angle between their fast axes. The interferometer is set in reflection in (b) with a programmable phase-only spatial light modulator (SLM) instead acting as the recombination MS. The SLM was placed in the back focal plane of the OL and used in reflection with a 50:50 beamsplitter. A QWP
preceeding the object converted the polarisation bases to linear. The light reflected back through the collection OL and exited the side port of the BS. The lens, $L_2$, acted in combination
with a second OL to image the object onto the CCD. A QWP in the reflection arm reverted the polarisation back to its circular components for projection onto the states \{H,V,D,A\} by rotating a HWP preceeding a horizontally orientated P. HWP: half waveplate, Obj.: Object, BS: beamsplitter (50:50), L: lens, CCD: charge-coupled device (camera).} 
\label{Fig2}
\end{figure} 

\begin{figure*}
\centering
\includegraphics[width=1\linewidth]{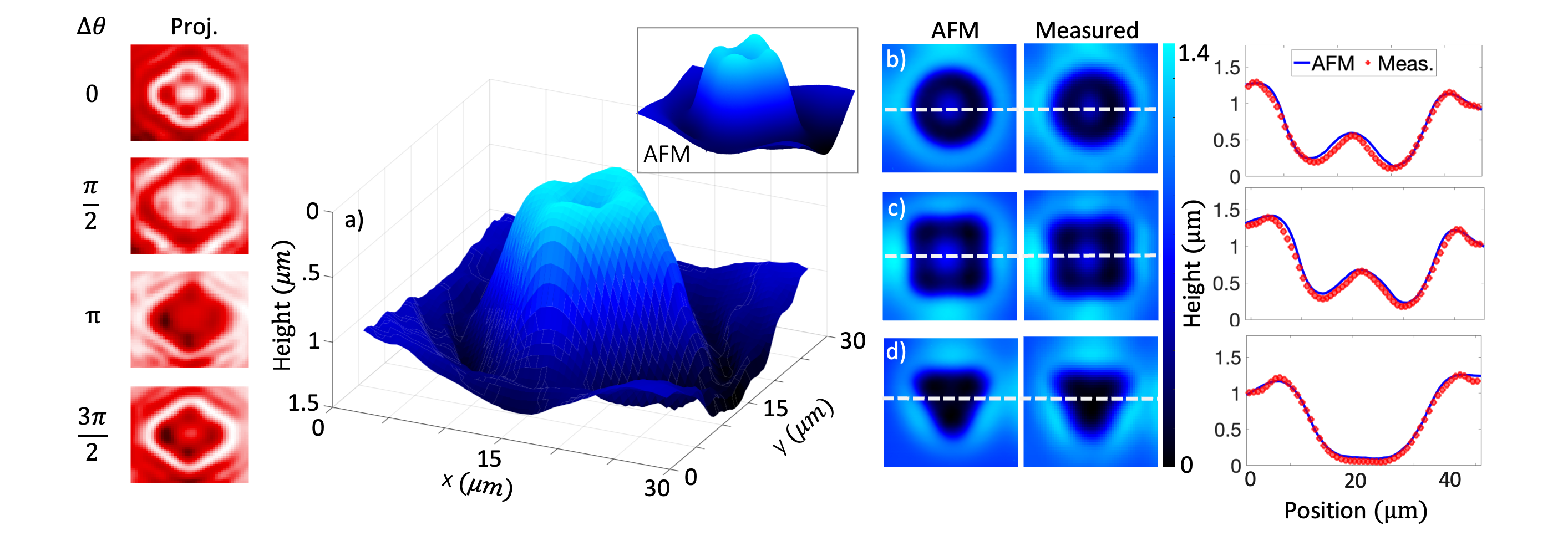}
\centering
\caption{\textbf{Quantitative phase imaging with a vortex interferometer.} Height measurements for samples with (a) diamond, (b) circular, (c) square and (d) triangular geometric shapes. Insets to the left of (a) show the intensities measured for intermodal phases $\Delta\theta$ = $\{0,\frac{\pi}{2},\pi,\frac{3\pi}{2}\}$ used to quantify the phase carried by the probe beam and reconstruct the changes in height across the sample. The AFM measurement is given in the topright corner. Height (\textit{z}-axis) is plotted in reverse to emphasise the sample's features. Rightmost graphs of (a), (b) and (c) show the measured height variations along the dotted lines on the 2D images in comparison with the AFM reference. Respectively, SLI measurements of the circle, square and triangular samples have RMSEs of (61 $\pm$ 5) nm, (71 $\pm$ 5) nm and (43 $\pm$ 2) nm from the AFM values. Error bars are smaller than the data points. Sample height is indicated by the false colormap to the right of the images with units also expressed in $\mu$m.}
\label{Fig3}
\end{figure*}

\begin{figure}
\centering
\includegraphics[width=1\linewidth]{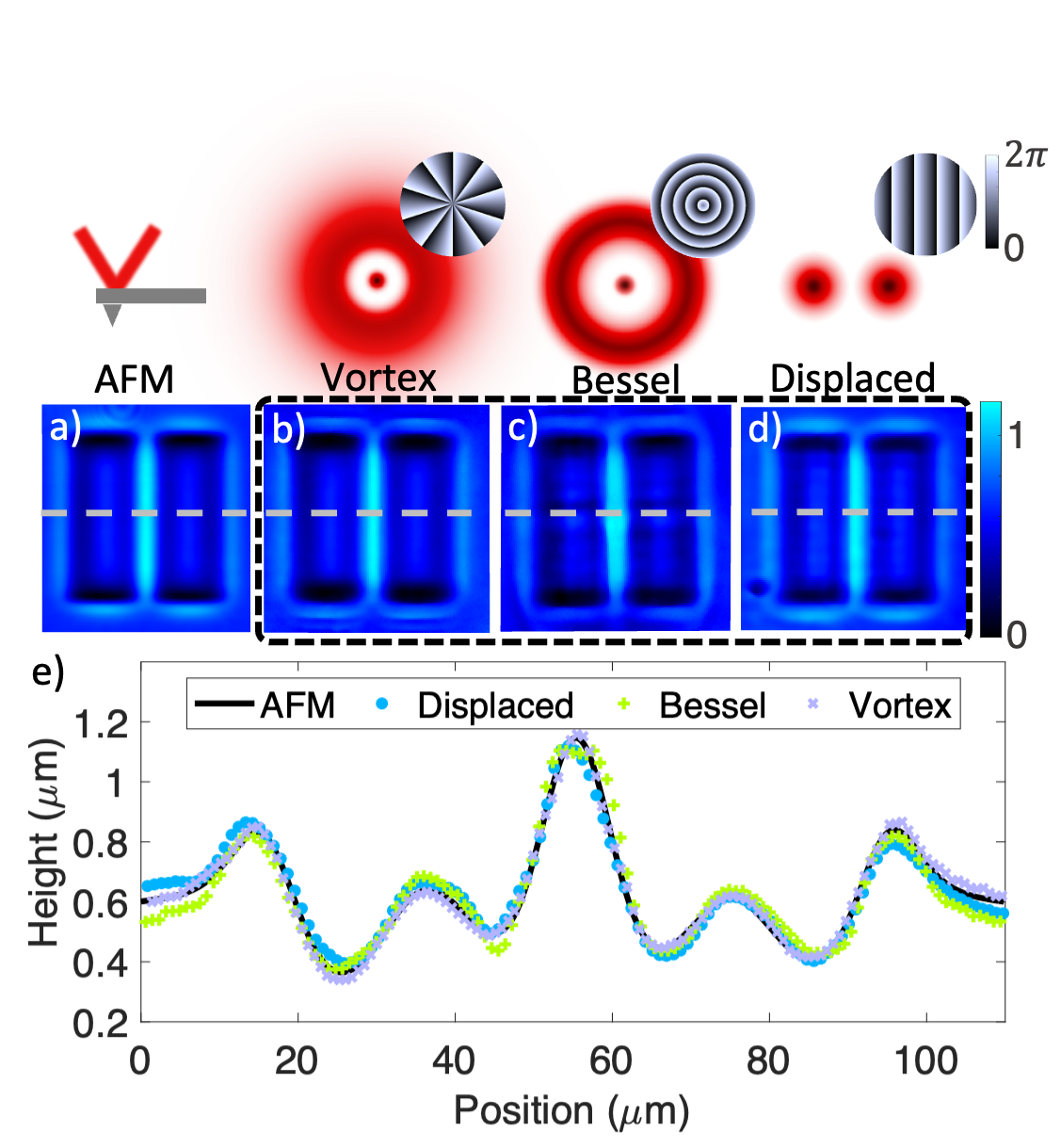}
\centering
\caption{\textbf{Quantitative phase imaging with different reference modes.} Comparison of $\mu m$ height variations measured for rectangular bar shaped samples using (a) an AFM reference and SLIs with (b) Vortex, (c) Bessel and (d) Displaced structured modes. Horizontal cross-sections of the variation in height along the grey dotted lines are plotted in (e) for the AFM (black solid line) as well as the Displaced (blue dots, RMSE = (32.7 $\pm$ 1.3) nm), Bessel (green crosses, RMSE = (45 $\pm$ 3) nm) and Vortex (purple x, RMSE = (18 $\pm$ 1) nm) interferometer measurements. Error bars are smaller than the data points. Height is indicated by the false blue colormap in units of $\mu$m.}

\label{Fig4}
\end{figure}

\begin{figure*}
\centering
\includegraphics[width=1\linewidth]{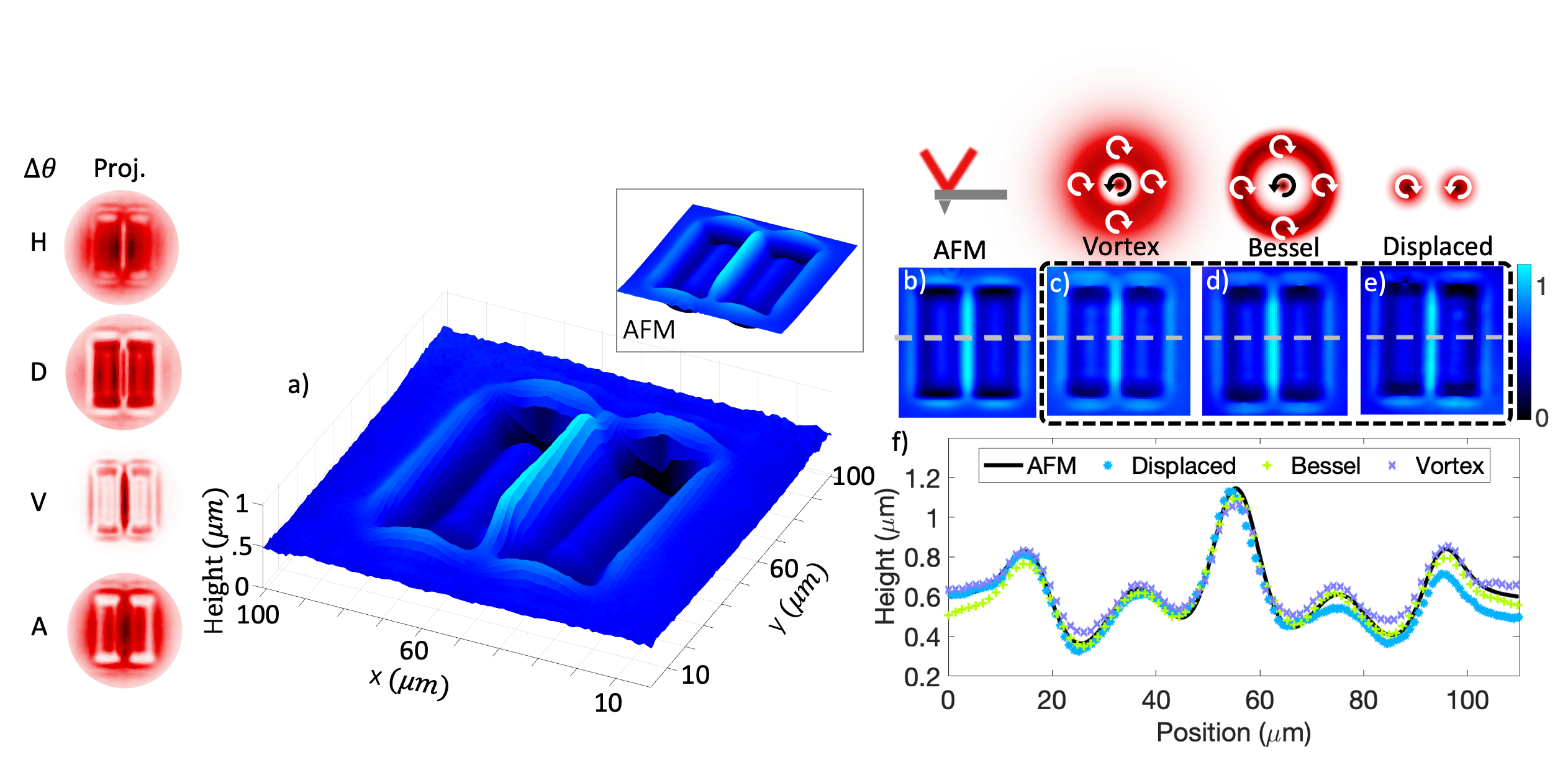}
\centering
\caption{\textbf{Quantitative phase imaging by mapping the phase object to polarisation.} Height measurements for transparent rectangular bars (a-e), where the imparted phase is mapped to polarisation in a vortex interferometer (a,c). Insets to the left of (a) show exemplary measured intensities from polarisation projections $\{H,D,V,A\}$ used for the height reconstruction (central 3D plot in (a) and 2D plot in (c)). Top-right corner 3D inset in (a) and 2D plot (b) depict the AFM measurement for comparison. Height measurements of the same sample using the (d) Bessel and (e) Displaced structured interferometers are also shown. Horizontal cross-sections of the variation in height along the grey dotted lines in (b-e) are plotted in (f) for AFM (black solid line), Displaced [blue dots, RMSE = (59 $\pm$ 4) nm], Bessel [green crosses, RMSE = (37 $\pm$ 3) nm] and Vortex [purple x, RMSE = (44 $\pm$ 2) nm] interferometers. Error bars are smaller than the data points. Height (in $\mu m$) is indicated by the false blue colormap. Circular arrows on the simulated interferometer structures in the object plane (above 2D plots) highlight the orthogonal polarisation states maintained when the reference and probe structures are recombined.}
\label{Fig5}
\end{figure*}
 
 \noindent \textbf{Retrieving phase with intensity mapping.} It is straightforward to show (see SI) that this transformation can be achieved by half-tuning the second MS and projecting this recombined field onto its RCP component, $\Psi_{R}(r,\phi) = \frac{1}{2}\left(e^{i Obj(x,y)}+e^{i\Delta\theta}\right)\mathcal{F}[G(r)]\hat{e}_R$. As illustrated in Fig. \ref{Fig1} (d), the output field is uniformly polarised with the object phase information modulating the probe intensity ($I_{\Delta\theta}(x,y) = |\Psi_{R}(r,\phi)|^{2}$), 
\begin{equation}
 \begin{split}
    I_{\Delta\theta}(x,y) = \frac{1}{2}[1+\cos(Obj(x,y)  - \Delta\theta)]|\mathcal{F}[G(r)]|^2,
    \end{split}
\end{equation}
in the camera plane.

Precise quantitative calculation of this phase typically requires four intensity measurements \cite{tyo2002design,zallat2006optimal} in the image plane of the object. Here, the intermodal phase can be controlled (details in Fig. \ref{Fig2}) to introduce the values $\Delta\theta = \{0, \frac{\pi}{2}, \pi, \frac{3\pi}{2}\}$ with respect to the probe. The phase retardation associated with the object is reconstructed as
\begin{equation}
    Obj(x,y) = \tan^{-1}\left(\frac{I_{\frac{\pi}{2}}(x,y)-I_{\frac{3\pi}{2}}(x,y)}{I_{0}(x,y)-I_{\pi}(x,y)}\right).
    \label{Eq.PhaseRecon}
\end{equation}
The same (Eq. \ref{Eq_split} - \ref{Eq.PhaseRecon}) holds true for other modes such as the Bessel and displaced beams depicted alongside the vortex in Fig. \ref{Fig1} (c) where the generation and recombination optics are easily replaced with ones inducing alternative phase structures.

To experimentally validate our framework, we quantitatively imaged in-house fabricated azopolymer samples with different topographies (see Methods and SI), and used atomic force microscopy (AFM) as a gold-standard benchmark to verify the accuracy of our results. We first realised a vortex interferometer by passing a RCP laser beam with a Gaussian spatial mode through two identical half-tuned $q$-plates with $q$ = 5 (Fig. \ref{Fig2} (a) setup). The samples were placed in the Fourier plane between the MSs (shown as Obj.) and centred on the Gaussian component as depicted in Fig. \ref{Fig1} (a,i). Afterwards, the light exiting the SLI was projected onto its RCP component with a quarter waveplate (QWP) and polariser before being imaged onto a CCD. 
Control of $\Delta\theta$ was easily achieved inside the SLI by inserting a pair of half-waveplates and rotating one of them (HWP$_{1,2}$, see Fig. \ref{Fig2} (a) for more detail). This introduced a delay between the orthogonal circularly polarised modes depending on the orientation of the half-wave plate’s fast axis.

Starting with a diamond-shaped sample in Fig. \ref{Fig3} (a), we show exemplary intensity measurements made in the image plane of the object (leftmost column) for various intermodal phases. Different spatial regions cycle through constructive (red) and destructive (white) interference related to the phase imparted by the sample as $\Delta\theta$ was changed. From Eq. \ref{Eq.PhaseRecon}, we quantified these phases at each pixel and estimated the local height (Methods, Eq. \ref{Eq.Height}), thereby obtaining a relative measurement of the sample topography. This is shown in the central three-dimensional plot of Fig. \ref{Fig3}. Qualitatively, our measurements agree very well with the AFM benchmarks (top right inset). To further illustrate the technique, we measured additional samples with circular, square and triangular relief geometries in Figure \ref{Fig3} (b), (c) and (d), respectively. False colormap images of the change in height measured across the samples are shown for both the AFM and our interferometer. 
Cross-sectional plots along the x-axis of the images (white dotted line), reveal excellent agreement between our interferometer (red dots) and the AFM (blue lines). Their root mean square errors (RMSE) all fall below 72 nm, while similarities remain above 95\% (see SI for more details). Small deviations in the local features are the result of experimental imperfections in the optical alignment.

 We next demonstrate the versatility of SLIs by designing interferometers with different modal structures.
Still keeping to half-tuned geometric-phase devices, the MSs (Fig. \ref{Fig2} (a)) were changed for ones imparting either a conical phase of $\Phi(x,y)= k_\gamma r$ (geometric-phase axicon \cite{ren2018micro} with slope $k = 6.67$ rad/mm) or a horizontal linear phase ramp, $\Phi(x,y)=\frac{2\pi x}{\Lambda}$ ($g$-plate \cite{d2020two} with spatial period $\Lambda$ = 75 $\mu$m). We label these interferometers Bessel and displaced, respectively, and show the transverse spatial structures (simulated) for each interferometer as they appear in the plane of the object (upper row of Fig. \ref{Fig4}) together with each MS's phase profile (top right corner). For each architecture, the topography of a sample with double rectangular bar geometry was measured (Fig. \ref{Fig4} (b), (c) and (d), respectively). Comparison with the AFM measurement (Fig. \ref{Fig4} (a)), shows visual agreement and consistency across all the interferometer types. Small variations in the defining edges of the shape can be attributed to aberrations in the imaging system and aperturing effects that reduce the collection of higher frequency components refracting from the sample. We confirm very close agreement in Fig. \ref{Fig4} (e) with a cross-sectional plot of the topography along the \textit{x}-axis of the 2D plots (dashed grey lines). Each of the height profiles closely follows the AFM measurement (solid black line) with RMSEs falling below 45 nm and similarities above 97\%. 

\noindent \textbf{Retrieving phase with polarisation mapping.}  Previously, the recombination MS restored the reference beam's polarisation along with the modal structure. However, SLIs offer the choice to map phase information in polarization rather than intensity. Fig.~\ref{Fig1} (e) illustrates that keeping the polarisation of the two modes orthogonal produces a spatially varying polarisation distribution in the camera plane that depends on the phase object,
\begin{equation}
    \Psi(r,\phi)\hat{e}(r,\phi) = \frac{1}{\sqrt{2}}\left(\hat{e}_L + e^{i Obj(x,y)}\hat{e}_R\right)\mathcal{F}[G(r)].
    \label{Eq. Pol}
\end{equation}
This way, the information can be encoded in and retrieved from the spatially-varying polarisation distribution. The equally weighted RCP and LCP components in Eq. \ref{Eq. Pol} form linear states so that the intermodal phases between the reference and probe (Eq. \ref{Eq.PhaseRecon}) can be set by projecting onto the horizontal (H), diagonal (D), antidiagonal (A) and vertical (V) polarisation components \cite{cui2022quantitative} of the recombined light outside the interferometer. 

This is readily achieved with devices such as SLMs which affect no change on the polarisation and modulate only the horizontally polarised component of light. Switching to the experimental configuration in Fig.
\ref{Fig2} (b), we replaced the recombination MS with an SLM. A QWP then transformed the structured (unstructured) mode's polarisation to horizontal (vertical) before it impinged onto the SLM where the inverse modal transformation was programmed to recombine the reference mode with the probe. Outside the interferometer, another QWP reverted the polarisation basis back to circular, and a HWP with a fixed polariser projected onto the linear polarisation components.

Using the same double-bar sample in this polarisation-encoding vortex interferometer, we show how the linear polarisation projections (H,D,V and A) control the intermodal phase [left of Fig. \ref{Fig5} (a)]. The reconstructed surface relief is shown in Fig. \ref{Fig5} (a) with the two-dimensional rendering of the same dataset also given in Fig. \ref{Fig5} (c). Panels Fig. \ref{Fig5} (d) and Fig. \ref{Fig5} (e) show the experimental results for the Bessel and displaced interferometer architectures, respectively. Horizontal cross-sections [Fig. \ref{Fig5} (f)] show good agreement with the AFM benchmark. Here, the similarities exceed 96\% and RMSEs are below 60 nm. It follows that these results are also in close agreement with those from the intensity encoding scheme of Fig. \ref{Fig4}.

\section*{Discussion and conclusion}

Each SLI inherits the properties of its underlying modes, offering substantial freedom in its design. The spatial symmetry, for instance, is set directly by the structured modes themselves. While we showed cylindrical (vortex and Bessel modes) and Cartesian (displaced mode) symmetries, the approach can be extended to many other tailored architectures. For instance, Ince-Gaussian \cite{bandres2004ince} and Airy beams \cite{efremidis2019airy} respectively provide elliptical and more complex Cartesian geometries. 
Beam size dependence is another factor, where both the displaced and Bessel modes carry a radial dependence in their phase functions. This requires care in the magnification used between the MSs or size-matching the hologram (see Methods). In contrast, vortex modes with a purely azimuthal phase structure exhibit no such size dependence  \cite{litvin2012azimuthal}. 
At the same time, SLI architectures support fine-tuning of modal parameters to suit the application. For example, $q$ and $k_{\gamma}$ directly govern the spatial separation from the probe and the ring size, providing a high degree of control. Moreover, modal features may offer the potential for enhanced performance, such as self-healing in Bessel modes \cite{shen2022self} and stability over long distances \cite{zhang2021generation}.

Our approach is also readily accessible across a range of devices, leveraging the well-established toolkits for generating and manipulating structured light \cite{forbes2020structured,forbes2021structured}. This shows that the concept is not tied to a specific platform, but can instead be adapted to the demands of the application. For example, tunable geometric-phase elements offer compact, wavelength-adjustable implementations \cite{slussarenko2011tunable}, metasurfaces provide high-resolution and asymmetric operations \cite{devlin2017arbitrary} as well as broadband functionality \cite{xia2025ultrabroadband}, and spatial light modulators offer dynamic, on-demand control of structured modes. Additionally, while we demonstrate our technique for an even split of power into the probe and reference modes, it is also possible to exploit and uneven ratio to compensate for losses and increase contrast when desirable.

Another interesting aspect is the implementation of the recombination step and the associated role of polarisation. Here we show that it determines whether phase is mapped to the intensity or to the polarisation state of the field. Whereas both approaches are effective, polarisation encoding offers additional benefits \cite{cui2022quantitative}, enabling a single-shot, real-time measurements with a polarisation-sensitive camera. In this setup, superpixels record all the projections simultaneously. Additionally, this can also facilitate measuring complex-amplitude objects where both amplitude and phase can be quantified at the same time.

Even though we implemented SLIs for imaging the phase of transmissive samples, this scheme can also be adapted for reflective ones. The working principle and procedure remain the same, with the generation optic also performing the recombination as light is reflected back through the setup (a folded interferometer). 
Interestingly, phase imaging with structured light has previously been restricted to edge enhancement of phase steps \cite{zangpo2023edge,gozali2017compact}, distinguishing elevation \cite{furhapter2005spiral} or computational analysis by decomposition into their bases \cite{rodriguez2022measurement} with exception to OAM where suitable gradients for computationally assisted phase reconstruction was found in a self-referencing configuration \cite{bernet2006quantitative,zeng2022optical}. Although grating-based techniques have been previously explored \cite{yasin2018path}, they represent a specific instance of the broader interferometric principle introduced here. In our framework, the structured modes themselves function as the interferometer arms, providing a compact and robust basis for both phase-shifting and polarization metrology.

In conclusion, we have introduced a route to common-path interferometry within a single optical beam by utilizing structured modes as interferometric arms. This architecture is inherently versatile, supporting a diverse range of modal structures and can be implemented using readily available dynamic and geometric phase devices. We demonstrate its performance through quantitative phase imaging, where our measurements across different SLI configurations exhibit an excellent agreement with atomic force microscopy benchmarks. Notably, by controlling the modal recombination, phase information is encoded into intensity or polarization states. The latter enables phase retrieval via external projective measurements, providing a clear path toward real-time single-shot imaging. This robust and accessible framework offers an approach that can be easily integrated into existing systems, ranging from manufacturing test beds to investigations of fundamental physical phenomena. While demonstrated here for a single application, we emphasise that our paradigm is fundamentally general and extends to a wide array of interferometric scenarios, offering a promising platform for high-precision sensing across materials science, biology, and industrial diagnostics. 
\section*{Methods}

\noindent \textbf{Experimental setup imaging variations}
The focal lengths in the experimental setup varied based on the configurations shown in Fig. \ref{Fig2} (a) and (b) of the main text. Within Fig. \ref{Fig2} (a), $f_1 = 400$ mm for the vortex-based interferometer, while $f_1 = 45$ mm (4x OL) for the other SLIs. Lens $L_2$ had a focal length of $f_2 = 200$ mm, which acted in combination with the OL. In Fig. \ref{Fig2} (b), $f_1 = 400$ mm for all the cases and the lens, $L_2$ was chosen to have a focal length of $f_2 = 100$ mm. \\

\noindent \textbf{Size-dependency of the recombination step}
We note that due to the radial dependence of the axicon plate and g-plate structures, the beam traversing identical ones for generation and reversal of the structured modes must have the same size so that the gradient of the conjugate phase across the beam is the same.  As a result, either the imaging lenses between the structuring elements should give no magnification or the parameters of second optic should be adjusted to compensate for a magnification factor between the conversion elements. We demonstrate both approaches. In the first experimental setup, Fig. \ref{Fig2} (a), the MSs were identical and the imaging optics chosen to have a magnification of 1 (OL with $f_1 = 45$ mm). This allowed the beam size in the back focal plane of the focusing and collection objectives to be same. 
Conversely, in the second experimental setup (Fig. \ref{Fig2} (b)), $L_1$ had a focal length of $f_1 = 400$ mm as the phase gradients could be pragmatically compensated for on the SLM (Meadowlark 1920x200 E-Series). The hologram was adjusted as $k_\gamma \rightarrow k_\gamma w_0/w_2$ for the Bessel mode and  $\Lambda \rightarrow \Lambda w_0/w_2$ for the displaced mode in order to compensate for the reduced beam size ($w_2$).
As vortex beams have no radial dependency in the phase function, there was no need to adjust for beam size in the experimental setup of Fig. \ref{Fig2} (a). As such, $L_1$ had a focal length of $f_1 = 400$ mm.  We further note that for optimal transformation of the structured mode back into the initial state, the recombination optic should be placed in the image plane of generation optic. This negates the need to account for additional propagation-related phases.\\

\noindent \textbf{Samples and topography retrieval.}
To test the system, we fabricated microscale topographic surface reliefs on azopolymer films acting as transparent samples with a homogeneous refractive index ($n = 1.61\pm0.02$ at $\lambda = 803$ nm), by means of holographic lithography \cite{oscurato2018nanoscopic,oscurato2022shapeshifting,reda2023reprogrammable} (see SI for more information).
The sample's topography was measured using atomic force microscopy (AFM) for comparison with the measurements from our SLI. Here, the sample thickness $h(x,y)$ imparts a local change in the phase of the probe beam, described by 
\begin{equation}
   \varphi(x,y) = h(x,y) \times \frac{2\pi(n - n_{0})}{\lambda},    \label{Eq.Height}
\end{equation}
where $n_0$ is the refractive index of air. Topographies of each sample were retrieved from our phase measurements (Eq. \ref{Eq.PhaseRecon}) at each pixel $(x,y)$ by calculating $h$ from Eq. \ref{Eq.Height}. \\

\noindent \textbf{Funding} The authors acknowledge the funding from the Italian Ministry of Research (MUR) through the PRIN 2022 project QNoRM and the PNRR  project PE0000023-NQSTI. We also acknowledge support by the ERC grant HyperMaSH, 101164874, funded by the European Union. Views and opinions expressed are however those of the authors only and do not necessarily reflect those of the European Union or the European Research Council Executive Agency. Neither the European Union nor the granting authority can be held responsible for them.  \\

\noindent \textbf{Acknowledgments} The authors would like to thank Prof. Lorenzo Marrucci for helpful and stimulating discussions.

\noindent \textbf{Disclosures} The authors declare no conflicts of interest.

\noindent \textbf{Data availability} Data underlying the results presented in this paper may be obtained from the authors upon reasonable request.

%\bibliography{refs.bib}

\newpage \clearpage

\setcounter{page}{1}
\setcounter{section}{0}
\setcounter{equation}{0}

\onecolumngrid
\begin{center}
    \textbf{\Large Supplementary Information: Designing interferometers within a single optical beam}

    \vspace{0.5 cm}

    Bereneice Sephton,$^1$ Rakhi Thomas,$^{1}$ Carlo Schiano,$^{1}$ Francesco Reda,$^1$ 
    I Komang Januariyasa,$^1$
    Filippo Cardano,$^1$ Bruno Piccirillo,$^{1}$
 Marcella Salvatore,$^{1}$ Stefano Luigi Oscurato,$^{1}$ Corrado de Lisio,$^1$ and Vincenzo D'Ambrosio$^1$

\vspace{0.5 cm}
    
$^1$Dipartimento di Fisica "E. Pancini", University of Naples Federico II, Naples, Italy

\end{center}

\vspace{0.5 cm}

\twocolumngrid

\section{Liquid crystal tuneable geometric phase devices} 

Our interferometers were structured and unstructured by exploiting the generation of polarisation-dependant geometric phase from a birefringent liquid crystal slab with electronically tunable retardation. Depending on the patterning of the optical axes across the device, a spatially varying phase is imparted to the incident light. Using Jones matrix formalism, the general operation of these conversion optics (MS) in the circular polarisation (CP) basis can be described by
\begin{equation}
    U(\delta,\alpha)_{MS} = \cos{(\frac{\delta}{2}})\begin{bmatrix}
        1& 0\\
        0& 1
    \end{bmatrix} +
    \sin{}(\frac{\delta}{2})\begin{bmatrix}
        0&ie^{-i2\alpha(x,y)}\\
        ie^{i2\alpha(x,y)}&
    \end{bmatrix},
\end{equation}
where $\alpha(x,y)$ refers to the local optical axis orientation in the transverse plane of Cartesian coordinates $(x,y)$ and $\delta$ is the retardation, tunable by applying a voltage.  

\textbf{Q-plates:} For vortex beams, the optical axis is made to vary azimuthally, where $\alpha = q\phi + \alpha_0$. The topological charge, $q$, specifies the number of rotations of the LC director in any closed counterclockwise path around the origin where $\phi = \tan{}^{-1}(y/x)$ is the azimuthal coordinate and $\alpha_0$ an offset angle. Aptly named the q-plate, we can describe its operation by substituting in $\alpha$ to give
\begin{equation}
    U(\delta,q\phi)_{MS} = \cos{}(\frac{\delta}{2})\begin{bmatrix}
        1& 0\\
        0& 1
    \end{bmatrix} +
    i\sin{}(\frac{\delta}{2})\begin{bmatrix}
        0&e^{-i2q\phi}\\
        e^{i2q\phi}&0
    \end{bmatrix},
\end{equation}
with $\alpha_0 = 0$ for convenience. When fully tuned ($\delta = \pi$) only the rightmost term remains. It is easily be seen that incidence with right circular polarised light ($\ket{R} = [1 ; 0]^T$),
\begin{equation}
    U(\pi,q\phi)_{MS}\ket{R} = 
    \begin{bmatrix}
        0&e^{-i2q\phi}\\
        e^{i2q\phi}&0
    \end{bmatrix}
    \begin{bmatrix}
        1\\
        0
    \end{bmatrix}
    = e^{i2q\phi}\begin{bmatrix}
        0\\
        1
    \end{bmatrix},
\end{equation}
generates orbital angular momentum (OAM) of $\ell\hbar = 2q\hbar$ per photon and the polarisation is flipped to left circular ($\ket{L} = [0;1]^T$). Similarly, incidence with $\ket{L} \rightarrow e^{-i2q\phi}\ket{R}$, where the OAM imparted is now negative. 
Generation of the co-propagating interferometer, however, requires an equal superposition of the initial Gaussian state ($\ell = 0$) as well a vortex. Half tuning the q-plate easily achieves this,
\begin{equation}
\begin{aligned}
    U(\pi/2,q\phi)_{MS}\ket{R} &= 
    \frac{1}{\sqrt{2}}\begin{bmatrix}
        1&ie^{-i2q\phi}\\
        ie^{i2q\phi}&1
    \end{bmatrix}
    \begin{bmatrix}
        1\\
        0
    \end{bmatrix} \\
    &= \frac{1}{\sqrt{2}}\begin{bmatrix}
        1\\
        0
    \end{bmatrix} + ie^{i2q\phi}\frac{1}{\sqrt{2}}\begin{bmatrix}
        0\\
        1
    \end{bmatrix} \\
    &=\frac{1}{\sqrt{2}}(\ket{R}+ ie^{i2q\phi}\ket{L}),
\end{aligned}
\end{equation}
with only half the light gaining OAM and flipping polarisation.

\textbf{Axicons:} To generate Bessel-Gauss modes, the optical axis is patterned to impart the phase of an axicon, with $\alpha = k_\gamma r + \alpha_0$. Here $r = \sqrt{x^2+y^2}$ and $k_\gamma = \frac{2\pi\sin{(\gamma)}}{\lambda}$ specifies the gradient of the conical phase extending from the origin. $\pi - \gamma$ represents the apex angle of the cone. Still assuming $\alpha_0=0$, the geometric-phase axicon here is thus characterised by the matrix, 
\begin{equation}
    U(\delta,k_\gamma r)_{MS} = \cos{}(\frac{\delta}{2})\begin{bmatrix}
        1& 0\\
        0& 1
    \end{bmatrix} +
    i\sin{}(\frac{\delta}{2})\begin{bmatrix}
        0&e^{-ik_\gamma r}\\
        e^{ik_\gamma r}&0
    \end{bmatrix},
\end{equation}
where right (left) CP results in the generation of a beam with positive (negative) conical phase, generating a Bessel mode that forms a well-defined ring in the farfield. Half tuning the optic so that $U(\delta,k_\gamma r)_{MS}\ket{R} \rightarrow \frac{1}{\sqrt{2}}(\ket{R}+ ie^{ik_\gamma r}\ket{L})$ generates the Bessel mode with orthogonal circular polarisation such that it forms a circular reference around the unconverted Gaussian mode. 

\textbf{G-Plates:} Our Displaced modes are generated by imparting a linear phase gradient to a percentage of the beam so that it spatially separates from the optical axis with propagation, while marking each with orthogonal polarisation. To do so, the optical axis takes on the orientation of a periodic structure for one direction, such as the $x$-axis, where $\alpha = \frac{\pi}{\Lambda}x + \alpha_0$. Here $\Lambda$ specifies the spatial period and $\alpha_0$ (which is zero in our case) remains an off-set angle. The device is then characterised by the matrix, 
\begin{equation}
    U(\delta,\frac{\pi x}{\Lambda})_{MS} = \cos{}(\frac{\delta}{2})\begin{bmatrix}
        1& 0\\
        0& 1
    \end{bmatrix} +
    i\sin{}(\frac{\delta}{2})\begin{bmatrix}
        0&e^{-i\frac{2\pi x}{\Lambda}}\\
        e^{i\frac{2\pi x}{\Lambda}}&0
    \end{bmatrix}.
\end{equation}
Fully tuned (second term), the plate performs the same flip in polarisation, while a imparting a phase gradient of $+(-)\frac{2\pi}{\Lambda}x$, causing the beam to propagate to the left (right) based on the incident circular polarisation, essentially rendering it a polarisation grating. By again half-tuning the optic such that $U(\delta,\frac{\pi x}{\Lambda})_{MS}\ket{R} \rightarrow \frac{1}{\sqrt{2}}(\ket{R}+ ie^{i\frac{2\pi x}{\Lambda}}\ket{L})$ only the reference (converted) beam propagates at an angle, allowing the unmodulated beam to easily interact with the sample at normal incidence. 

\section{Sample Fabrication}
The transparent phase-modulating samples measured in the SLIs were fabricated through a photopatterning process based on light-induced surface modulation in an azopolymer thin film \cite{oscurato2018nanoscopic}. The photoresponsive material was synthesized via radical polymerization of (E)-2-(4-((4-methoxyphenyl)diazenyl)phenoxy)ethyl acrylate. The polymer was dissolved in 1,1,2,2-tetrachloroethane at a concentration of 140 mg/mL, filtered through a 0.2 $\mu$m PTFE membrane, and spin-coated onto clean glass coverslips at 300 rpm for 4 minutes. This yielded homogeneous films with a thickness of approximately 1.7 $\mu$m.

The surface patterning was accomplished through digital holographic lithography, implemented by a reflective phase-only SLM (Holoeye Pluto 2.1) \cite{oscurato2022shapeshifting,reda2023reprogrammable}. An expanded laser beam with a wavelength of 491 nm was modulated using phase maps (kinoforms) displayed on the SLM. The kinoforms were computed by an iterative Fourier transform algorithm, employing a digital representation of the intended sample topography. Before incident on the azopolymer, the modulated beam was propagated through a 4-f optical system to the back focal plane of a 50× long-working-distance objective. This process enabled the conversion of kinoform information into optical intensity distributions, thereby reproducing the target light pattern on the photo-responsive film surface. Pattern inscription was performed with an average power density of 6.5 W/cm$^2$ and circular polarization, to ensure isotropic material response. To enhance the patterning dynamics and prevent photo-induced birefringence, a circularly polarized assisting beam (0.6 W/cm$^2$) with a wavelength of 405 nm  was simultaneously directed from the substrate side during exposure. When exposed to the patterned intensity, the azopolymer underwent directional mass migration at the microscale in response to the light distribution. This process lead to material accumulation and depletion in regions of high and low optical intensity, respectively. Consequently, the projected intensity pattern was transformed into a stable and permanent surface corrugation that accurately reproduced the target geometry. It is important to note that the surface topography was ready for immediate use following the inscription process, obviating any need for chemical or physical post-processing. The holographic exposure time was 20 and 50 seconds for the geometric (diamond, circular, square, and triangular) and dual-bar patterns, respectively.

The topographies of the fabricated azopolymer reliefs were characterized by Atomic Force Microscopy (AFM). Measurements were performed using a WITec Alpha RS300 microscope operating in tapping mode and Arrow FM cantilevers (resonance frequency of 75 kHz, nominal spring constant of 2.8 N/m, and tip radius of $\sim$10 nm).

\section{Half-tuned geometric phase plates for interfering the reference and probe} 

A salient point for using structured light to form an interferometer is how the reference is recombined with the probe. To reverse the initial creation step, an asymmetric operation is required in that both the spatial structure and polarisation of the reference is reverted, while the probe is unaltered. This can easily be achieved by enacting the same operation with an identical half-tuned geometric phase plate. However, as the incident light now has both CP components, $\ket{\Psi} = e^{i2\alpha(x,y)}\ket{L} + e^{iObj(u,v)}\ket{R}$, the tuned element works to also modulate the probe containing the object phase,
\begin{align}
        U(\frac{\pi}{2},\alpha)_{MS}\ket{\Psi} &= 
    \frac{1}{2}\begin{bmatrix}
        1&ie^{-i2\alpha(x,y)}\\
        ie^{i2\alpha(x,y)}&1
    \end{bmatrix}\begin{bmatrix}
        e^{iObj(u,v)}\\
        ie^{i2\alpha(x,y)}
    \end{bmatrix} \\
    &= 
    \frac{1}{2}\begin{bmatrix}
        e^{iObj(u,v)} - e^{-i2\alpha(x,y)}e^{i2\alpha(x,y)}\\
        ie^{i2\alpha(x,y)}e^{iObj(u,v)} + ie^{i2\alpha(x,y)}
    \end{bmatrix},
\end{align}
which is shown for the LCP component in the second entry. We omit the amplitude structures here for convenience and $(u,v)$ represents the Fourier transform related coordinates of $(x,y)$. If we consider only the RCP component however [$(e^{iObj(u,v)} - 1)\ket{R}$], it is clear that the reference phase structure ($e^{i2\alpha(x,y)}$) has been eliminated so that solely the phase of the object is present and can thus be measured. Projection onto this component is achieved with a quarter waveplate rotated 45$^\text{o}$ followed by a horizontally orientated polariser.

\section{Data analysis} 
Agreement between the extracted AFM reference and measured data profiles were analysed by interpolating each data set to the same sized vector and calculating both the root square mean error (RMSE),
\begin{equation}
    \text{RMSE} = \sqrt{\frac{\sum_{x}[h_\text{AFM}(x)- h_\text{SLI}(x)]^{2}}{N}}
\end{equation}
and similarity (S),
\begin{equation}
    \text{S} = 1- \frac{\sum_{x}\left|h_\text{AFM} (x) - h_\text{SLI}(x)\right|}{\sum_{x}[h_\text{AFM}(x)] + \sum_{x}[h_\text{SLI}(x)]}
\end{equation}
across all the elements. Here $x$ refers to the position, $N$ is the length of each dataset, $h_\text{AFM}$ is the AFM measured height and $h_\text{SLI}$ is the height derived from the structured light interferometers. Table \ref{tab:Comp} summarises these values for each profile shown in the main text.
\begin{table}[H]
    \centering
    \caption{Comparative analysis of extracted profiles between the AFM and SLI measured topographies. Amp. = Amplitude; Pol. = Polarisation}
    \begin{tabular}{|c|c|c|c|c|c|}
    \hline
        \textbf{Fig.}     & \textbf{Sample}   &   \textbf{SLI}    & \textbf{Mapping} & \textbf{RMSE (nm)}  & \textbf{$S$ (\%)} \\
    \hline
        3 (b) & Circle   &  Vortex  & Amp.     & 61 $\pm$ 5        & 96.3 $\pm$ 2.1 \\
    \hline
        3 (c) & Square   &  Vortex  & Amp.     & 71 $\pm$ 5        & 96.0 $\pm$ 1.6 \\
    \hline
        3 (d) & Triangle &  Vortex  & Amp.     & 43 $\pm$ 2        & 97.1 $\pm$ 1.8 \\
    \hline
        4 (b) & Bars     &  Vortex  & Amp.     & 18 $\pm$ 1         & 98.9 $\pm$ 0.9 \\
    \hline
        4 (c) & Bars     &  Bessel  & Amp.     & 45 $\pm$ 3          & 97.1 $\pm$ 1.5 \\
    \hline
        4 (d) & Bars     &  Displaced  & Amp.  & 32.7 $\pm$ 1.3          & 97.7 $\pm$ 1.3 \\
    \hline
        5 (c) & Bars     &  Vortex  & Pol.  & 44 $\pm$ 2          & 97.1 $\pm$ 1.6 \\
    \hline
        5 (d) & Bars     &  Bessel  & Pol.  & 37 $\pm$ 3          & 97.6 $\pm$ 1.6 \\
    \hline
        5 (e) & Bars     &  Displaced  & Pol.  & 59 $\pm$ 4       & 96.4 $\pm$ 1.6 \\
    \hline
    \end{tabular}
    \label{tab:Comp}
\end{table}
\end{document}